\documentclass[12pt]{article}
\usepackage[colorlinks,linkcolor=blue,citecolor=blue,urlcolor=blue,bookmarks,bookmarksnumbered]{hyperref}
\usepackage{amsmath,amssymb,amsfonts}
\usepackage[paper=letterpaper,margin=1.0in]{geometry}
\usepackage{graphicx,xcolor}
\usepackage{cite}

\newcommand{\be}{\begin{equation}}
\newcommand{\ee}{\end{equation}}
\newcommand{\bea}{\begin{eqnarray}}
\newcommand{\eea}{\end{eqnarray}}
\newcommand{\ba}{\begin{array}}
\newcommand{\ea}{\end{array}}
\newcommand{\ben}{\begin{enumerate}}
\newcommand{\een}{\end{enumerate}}
\newcommand{\bi}{\begin{itemize}}
\newcommand{\ei}{\end{itemize}}
\newcommand{\bc}{\begin{center}}
\newcommand{\ec}{\end{center}}
\newcommand{\bfig}{\begin{figure}}
\newcommand{\efig}{\end{figure}}
\newcommand{\bq}{\begin{quotation}}
\newcommand{\eq}{\end{quotation}}
\newcommand{\bt}{\begin{table}}
\newcommand{\et}{\end{table}}
\newcommand{\btab}{\begin{tabular}}
\newcommand{\etab}{\end{tabular}}
\newcommand{\bs}{\begin{slide}}
\newcommand{\es}{\end{slide}}

\newcommand{\pa}{\partial}

\newcommand{\IR}{\mathbb{R}}

\newcommand{\X}{\mathbb{X}}
\newcommand{\IP}{\mathbb{P}}

\def\pa{\partial}

\def\s{\sigma}

\let\a=\alpha

\let\ba=\overline

\def\e{\epsilon}

\let\l=\lambda
\let\L=\Lambda

\def\hx{\mathord{\hat x}}
\def\tx{\mathord{\tilde x}}
\def\htx{\mathord{\hat{\tilde x}}}

\let\w=\omega

\def\IR{\relax\leavevmode{\rm I\kern-.18em R}}
\def\ZZ{\relax\leavevmode
       \ifmmode\mathchoice
       {\hbox{\sf Z\kern-.4em Z}}
       {\hbox{\sf Z\kern-.4em Z}}
       {\lower.9pt\hbox{\scriptsize\sf Z\kern-.36em Z}}
       {\lower1.2pt\hbox{\tiny\sf Z\kern-.36em Z}}
       \else{\sf Z\kern-.4em Z}\fi}
\def\RR{\relax\leavevmode
       \ifmmode\mathchoice
       {\hbox{\sf R\kern-.4em R}}
       {\hbox{\sf R\kern-.4em R}}
       {\lower.9pt\hbox{\scriptsize\sf R\kern-.36em R}}
       {\lower1.2pt\hbox{\tiny\sf R\kern-.36em R}}
       \else{\sf R\kern-.4em R}\fi}

\def\resetby#1#2{\@addtoreset{#2}{#1}}
\def\seceq{\@addtoreset{equation}{section}
              \def\theequation{\thesection.\arabic{equation}}}

\def\Label#1{\label{#1}%
                \smash{\hbox to0pt{\raise1ex\hbox{\tiny[#1]}\hss}}}
\def\noLabels{\let\Label=\label}

\def\TeV{\text{T\kern0pte\kern-1ptV}}

\begin{document}
\thispagestyle{empty}
{\footnotesize
${}$
}

\bc

\vskip 1.0cm
\centerline{\Large \bf On Dark Energy and Quantum Gravity}
\vskip 0.5cm
\vskip 1.0cm

\renewcommand{\thefootnote}{\fnsymbol{footnote}}

\centerline{{\bf
Per Berglund${}^{1}$\footnote{\tt per.berglund@unh.edu},
Tristan H{\"u}bsch${}^{2}$\footnote{\tt thubsch@howard.edu}
and
Djordje Mini{\'c}${}^{3}$\footnote{\tt dminic@vt.edu}
}}

\vskip 0.5cm

{\footnotesize\it
${}^1$Department of Physics and Astronomy, University of New Hampshire, Durham, NH 03824, U.S.A. \\
${}^2$Department of Physics and Astronomy, Howard University, Washington, D.C.  20059, U.S.A. \\
${}^3$Department  of Physics, Virginia Tech, Blacksburg, VA 24061, U.S.A. \\
${}$ \\
}
\ec

\vskip 1.0cm

\begin{abstract}\unskip\noindent
\begin{quotation}\noindent
 Realizing dark energy and the observed de Sitter spacetime in quantum gravity has proven to be obstructed in most every usual approach. 
 We argue that additional degrees of freedom of the left- and right-movers in string theory and 
 a resulting doubled, non-commutatively generalized geometric formulation thereof can lead to an effective model of dark energy consistent with de Sitter spacetime.
 In this approach, the curvature of the canonically conjugate dual space provides for the dark energy inducing a positive cosmological constant in the observed spacetime, whereas the size of the  above dual space is the gravitational constant in the same observed de Sitter spacetime. As a hallmark relation owing to a unique feature of string theory which relates short distances to long distances, the cosmological constant scale, the Planck scale, and the effective \TeV-sized particle physics scale must satisfy a see-saw-like formula---precisely the generic prediction of certain stringy cosmic brane type models.
\end{quotation}
\end{abstract}

\vspace{1cm}

This essay received an honorable mention in the Gravity Research Foundation 2019

Awards for Essays on Gravitation competition



\renewcommand{\thefootnote}{\arabic{footnote}}

\newpage
\setcounter{page}{1}
\pagestyle{plain}

Whether or not asymptotically de Sitter space can be found in a consistent theory of quantum gravity and matter including string theory~\cite{Polchinski:1998rq} has captured much attention ever since the dramatic discovery of dark energy in the late 1990s~\cite{Riess:1998cb, Perlmutter:1998np}. (For the most recent measurements of the Hubble constant and the {\em\/associated discrepancies\/}, see~\cite{Riess:2016jrr, Riess:2018byc, Aghanim:2018eyx}; see below.)
 The existence and realization of de Sitter space as a solution in string theory is still considered and open question~\cite{Danielsson:2018ztv}, and the interest in this fundamental issue has been recently reignited in~\cite{Obied:2018sgi, Agrawal:2018own}; see also~\cite{Andriot:2019wrs,Palti:2019pca}. 
 We argue that  one can successfully address the problem of dark energy and the observed de Sitter spacetime in a generic, non-commutative generalized geometric phase-space formulation of string theory.
 Essentially, the curvature and size of the canonically conjugate dual space are the cosmological  and gravitational constants in the observed spacetime, respectively.
Furthermore, the three scales associated with:
 ({\small\bf1})~the cosmological constant,
 ({\small\bf2})~the Planck units and
 ({\small\bf3})~the effective particle physics,
are related by a see-saw-like formula via T-duality. This is a hallmark feature of string theory that relates reciprocally short and long distances, and precisely the generic prediction of certain toy models~\cite{rBHM7}.

In general, any theory of quantum gravity and matter 
is expected to produce the following low energy effective action valid at long distances of the observed accelerated universe (focusing on the relevant $3{+}1$-dimensional case~\cite{rAE-GR,rAE-GRcc}, in the ${+}\,{-}\,{-}\,{-}$ signature):
\be
S_{\text{eff}} = - \int d^4 x \sqrt{-g} \Big(\frac{1}{8 \pi G} \L + \frac{1}{16 \pi G} R + {\cal O}(R^2)
\Big),
 \label{e:Seff}
\ee
where string theory~\cite{
rF79a} introduces the ${\cal O}(R^2)$ correction terms.
However, it has proven difficult to produce the observed positive cosmological constant $\L$ within such a framework~\cite{Obied:2018sgi, Agrawal:2018own}. 

The generalized geometric formulation of string theory we have in mind has been recently discussed in~\cite{
Freidel:2017xsi, 
Freidel:2017nhg}, 
and derives from the underlying {\it chiral} world-sheet Hamilton's action for the strings~\cite{rWRH-Dyn1,rWRH-Dyn2}:
\be
S_{\text{string}}=  \frac{1}{4\pi}\int_{\Sigma}d^2\s
 \Big(\pa_{\tau}{\X}^{A} (\eta_{AB}+\w_{AB})\,\IP^{B} -  
  \IP^{A}  H_{\!AB}\,\IP^{B}\Big) ,
\label{e:MSA}
\ee 
where $\X^A$ combine the sum ($x^a$) and the difference ($\tx_a$) of the left- and right-movers on the string, and $\IP^A\,{=}\,\pa_\s\X^A$ are closely related to the (chiral) generalized momenta.
The mutually compatible dynamical fields $\w_{AB},\eta_{AB}$ and $H_{AB}$ are:
 ({\small\bf1})~the antisymmetric symplectic structure,
 ({\small\bf2})~the symmetric polarization metric $\eta_{AB}$ and
 ({\small\bf3})~the doubled symmetric metric $H_{\!AB}$, respectively.
This new framework for 
string theory based on a quantum space-time captures the essential quantum non-locality of any quantum theory~\cite{Freidel:2016pls}. 
 Also, $\w_{AB}$ governs the Hilbert structure of a quantum theory, which is usually ignored in the standard spacetime interpretation of string theory~\cite{Polchinski:1998rq}, whereas the usual Kalb-Ramond ($B_{\mu \nu}$) field is associated with the symplectic structure $\w_{AB}$, rather than the doubled metric~\cite{
 Freidel:2017nhg}.

In fact, quantization renders the doubled ``phase-space'' operators $\hat{\X}^A=(\hx^a/\l, \htx_b/\l)$ inherently non-commutative~\cite{rBHJ-OnQM2}, inducing in particular\footnote{In standard interpretations of string theory, the symmetric combinations of the left- and the right-movers, $x^a=x^a_L+x^a_R$, are identified with the target-spacetime coordinates. By effectively neglecting $\w_{AB}$ and imposing $[x^a,\tx_b]=0$, the canonically conjugate dual space, spanned by the difference of  the left- and right-movers $\tx_a=x^a_L-x^a_R$,  is often completely omitted.}~\cite{
Freidel:2017nhg}:
\bea
[ \hat{\X}^a, \hat{\X}^b] = i \w^{AB}:\qquad
[\hx^a,\hx^b]=0,\quad 
[\hx^a,\htx_b]=2\pi i\l^2 \delta^a{}_b,\quad  
[\htx_a,\htx_b]= 0,
\label{e:CnCR}
\eea
where $\l$ denotes the fundamental length scale, such as the Planck scale,
so that $\e=1/\l$ gives the corresponding fundamental energy scale.
This was found by examining the simplest example of the canonical free string compactified on a circle, in an intrinsically T-duality covariant formulation of the Polyakov string. Full spacetime covariance is maintained in this description and the string tension is naturally the ratio of the fundamental length and energy scales, $\a' = \l/{\e}$.
 This fundamental non-locality is independently confirmed by examining the algebra of vertex operators in the 2d CFT of a free string compactified on a circle~\cite{
 Freidel:2017xsi, 
 Freidel:2017nhg}.

The integration of the foundational concepts in physics, Hamilton's principle~\cite{rWRH-Dyn1,rWRH-Dyn2}, general relativity~\cite{rAE-GR,rAE-GRcc} and quantization~\cite{rBHJ-OnQM2} has fascinating physical consequences, some readily seen by exploring the chiral $\{\hx^a, \htx_b\}$-system:
 The non-trivial commutators~\eqref{e:CnCR} imply the corresponding complementary indeterminacy relation, $ \Delta {x}^a \Delta {\tx}_b\,{\sim}\,\l^2 \delta^a_b$, where the $\hx^a$ coordinate operators may be associated with short-distance (UV) ``spacetime,'' while the $\htx_b$ span reciprocally long (IR) distances.
It then naturally follows that all local effective fields must be regarded a priori as 
{\em\/bi-local\/} $\phi(x, \tx)$~\cite{Freidel:2016pls}, and subject to~\eqref{e:CnCR}, and therefore inherently non-local in the conventional $x^a$-spacetime. Such non-commutative field theories~\cite{Douglas:2001ba, Szabo:2001kg, Grosse:2004yu} generically display a mixing between the UV and IR physics. To have a well-defined continuum limit one has to appeal to a double-scale renormalization group (RG) and the self-dual fixed points~\cite{Grosse:2004yu,Freidel:2017xsi}. This implies that the effective field theory scale at low energies is the geometric mean of the UV and the IR scale. Such a double RG flow also leads to a world-sheet Lorentz invariant formulation of~\eqref{e:MSA} 
and at the relevant self-dual fixed point. By involving both parts of the phase-space, string theory possesses intrinsic
non-commutativity between $x$ and $\tilde{x}$ while the $x$'s still commute among themselves. This in turn connects the foundations of string theory to the deep results from non-commutative quantum field theory.


The so generalized geometric formulation of string theory discussed above provides for an effective description of dark energy that is consistent with a de Sitter spacetime due to its chirally doubled realization of the target space and the non-commutative structure in~\eqref{e:CnCR}. To this end, note that the natural stringy effective action on the doubled spacetime in terms of the coordinates $(x^a,\tx^a)$ takes the form:
\be
   S_{\text{eff}}^{\textit{nc}}
  =\int \text{Tr} \sqrt{g(x,\tx)}\, \big[R(x,\tx)+\dots\big],
    \qquad \text{with}~~[x, \tx] = i \l^2,
\label{e:ncEH}
\ee
where the ellipses denote higher-order curvature terms induced by string theory~\eqref{e:Seff}.  Owing to~\eqref{e:CnCR}, this $S_{\text{eff}}^{nc}$ clearly expands into a host of several terms, which upon $\tx$-integration and from the $x$-space vantage point gives rise to interactions that can lead to {\em\/various forms\/} of dark energy. Thus, these effects may provide for a way of addressing the recent conflicting measurements of the Hubble constant~\cite{Riess:2016jrr, Riess:2018byc, Aghanim:2018eyx}.

To lowest order the expansion of $S_{\text{eff}}^{\textit{nc}}$ takes the form:
\be
S_d = - \int \sqrt{-g(x)}  \sqrt{-\tilde{g}(\tx)} [R(x) + \tilde{R}(\tx)],
\label{e:TsSd}
\ee
a result which first was obtained almost three decades ago,  effectively setting $\w_{AB}\to0$ in~\eqref{e:CnCR} by assuming that $[\hat x^a,\htx]=0$~\cite{
Tseytlin:1990hn}. In this limit, the $\tx$-integration in the first term of~\eqref{e:CnCR} defines the gravitational constant $G_N$, and in the second term produces a {\it positive} cosmological constant constant $\L>0$. It also follows that the weakness of gravity is determined by the size of the canonically conjugate dual space, while the smallness of the cosmological constant is given by its curvature.
 However, these results from the commutative limit are not stable under loop corrections, which has been addressed in the recent work of Kaloper and Padilla (called the sequester mechanism) who also extended these results to loops of arbitrary order, in the effective field theory~\cite{Kaloper:2014dqa}.

The intrinsic non-commutativity of the zero modes $x$ and $\tx$~\eqref{e:CnCR} corrects these results in several ways. In particular, it is natural to ask whether the non-zero $\lambda$ in~\eqref{e:CnCR}    stabilizes the cosmological constant.
 The fully non-commutative analysis is intricate, but an encouraging indication emerges as follows:
 By simplifying to conformal metrics, $g_{\mu \nu} = \phi^2 \delta_{\mu \nu}$, the action~\eqref{e:ncEH}--\eqref{e:TsSd} produces a non-commutative $\L \phi^4$ theory.
 Unlike the theory's commutative limit, the beautiful results of Grosse and Wulkenhaar~\cite{Grosse:2012uv,Grosse:2014nza} demonstrate the non-perturbative solvability of the above non-commutative $\L \phi^4$ theory, explicitly showing the finite renormalization of $\L$ in terms of the bare coupling.
 At least in this highly simplified, conformal degree limit, non-commutativity thus can afford a small, radiatively and perhaps even non-perturbatively stable cosmological constant for the non-commutative form of the ``doubled'' effective action.

Finding de Sitter spacetime within string theory 
has been an on-going quest over the past two decades; see~\cite{Danielsson:2018ztv} for an excellent recent review, with an extensive list of references.
Among the vast number of various constructions we have developed a {\em\/discretuum\/} of toy models, see~\cite{
rBHM5, rBHM6} and references therein, that turn out to naturally capture several of the features of the above non-commutatively generalized phase-space reformulation of quantum gravity~\cite{rBHM7}. 
One of the essential features of our toy model is 
S-duality (the relation between the weak and strong coupling),
which is built in the SL$(2;\mathbb{Z})$ monodromy properties of our axion-dilaton models.
In generalizations where various moduli fields replace the axion-dilaton system,
this directly implies T-duality (the relation between the short and long distances),
which is in turn covariantly realized in the phase space approach.
Here T-duality maps $\hx$ into $\htx$, and vice versa,
and thus a covariant representation calls for a phase space formulation 
that involves both $\hx$ and $\htx$.
The see-saw formula is a hallmark of such a covariant phase space formulation, and, not surprisingly, it can be explicitly derived in the 
context of our toy models.
  The overall effect is thus closely related to the old observation of Witten~\cite{Witten:1994cga}, that supercharges need not be globally defined in the presence of conical defects, and the mass splitting between superpartners is controlled by the strength of the conical defect; for the corresponding four-dimensional generalization and relation to the cosmological constant, see~\cite{Jejjala:2002we,Becker:1995sp}.

Models in this class naturally produce a see-saw formula for the cosmological constant: 
\be
M_\L\,{\sim}\,M^2/M_P, 
\label{e:seesaw}
\ee
relating the mass scales of the vacuum energy/cosmological constant ($M_\L$), particle physics ($M$), and the Planck scale ($M_P$), respectively.
Specifically, the see-saw formula~\eqref{e:seesaw} follows from very particular geometric properties which relate both the volume and
the curvature of the space to the string length scale~\cite{rBHM5}.
 Identifying $M_\L$ and $M_P$ as the IR and UV cut-offs, respectively, the RG flow (in this non-commutative version) identifies a self-dual fixed point~\cite{Douglas:2001ba, Szabo:2001kg, Grosse:2004yu}. Given that the phase-space formulation~\cite{
 Freidel:2017xsi, 
 Freidel:2017nhg} is a T-duality covariant description of string theory, this naturally relates $M_P\,{\to}\,M^2/M_P$ under T-duality.
The prediction of our models ~\cite{rBHM6, rBHM7} $M_\L\,{\sim}\,M^2/M_P$ then satisfies these conditions, with $M_P\,{\sim}\,\e$ the fundamental energy scale corresponding to the fundamental length $\l$.
 This produces the well-known formula for the observed dark energy scale, provided $M$ is a \TeV\ scale. In addition, this illustrates one of the main points, namely their novel realization of dark energy and de Sitter spacetime, in both the phase-space formulation of string theory and the toy-models in this class~\cite{rBHM5,rBHM7}.

In conclusion, as we have argued in this essay, the doubled, non-commutative generalized geometric formulation of string theory leads naturally to a positive cosmological constant. Essentially, the curvature of the dual space is the cosmological constant in the observed spacetime, and the size of the dual space is the gravitational constant in the same observed spacetime. Also, the three scales associated with the effective particle physics (\TeV), the cosmological constant and the gravitational constant (Planck energy), are naturally arranged via a see-saw formula.
To see whether this proposal is consistent with other observations,
 one could look for cosmological signatures of intrinsic non-commutativity of string theory with generalized (Born) geometry. These cosmological signatures can have both UV and IR guises.
 In particular, one could look for non-local effects at the largest possible (Hubble) scale. A preliminary study of such non-local effects associated with a one-parameter vacuum structure of de Sitter space was discussed in~\cite{Kaloper:2002uj, Danielsson:2002kx, deBoer:2004nd}. Similarly, effective non-commutativity in cosmology was discussed in~\cite{Brandenberger:2002nq} and the minimal length in~\cite{
 Brandenberger:2006xi}, and references therein. 
 However, such results should be re-examined from the point of view of an effective non-commutative description of quantum gravity in the context of the phase-space formulation of string theory. Finally, our discussion naturally relates to
the observationally supported proposal for dark matter quanta that are sensitive to dark energy~\cite{
Edmonds:2017zhg}. 

\noindent
{\bf Acknowledgments:} 
DM thanks Laurent Freidel and Rob Leigh for numerous insightful discussions over many years on the topic of quantum gravity and string theory, and Vijay Balasubramanian, Petr Ho{\v{r}}ava and Jan de Boer for many conversations on de Sitter space and string theory.
 PB would like to thank the CERN Theory Group for their hospitality over the past several years.
 TH is grateful to the Department of Physics, University of Maryland, College Park MD, and the Physics Department of the Faculty of Natural Sciences of the University of Novi Sad, Serbia, for the recurring hospitality and resources.
 The work of DM is supported in part by the Julian Schwinger Foundation.

\footnotesize\baselineskip=2.5ex
%
%

\end{document}